\documentclass[aps,prl,twocolumn,showpacs,groupedaddress]{revtex4}
\usepackage{graphics}
\begin{document}
\title{Modulated Vortex Lattice in High Fields and Gap Nodes}

\author{Ryusuke Ikeda$^1$ and Hiroto Adachi$^2$}
\affiliation{
$^1$Department of Physics, Kyoto University, Kyoto 606-8502, Japan
\\ $^2$Department of Physics, Okayama University, Okayama 700-8530, Japan}

\date{\today}

\begin{abstract}

The mean field vortex phase diagram of a quasi two dimensional superconductor with a nodal $d$-wave pairing and with strong Pauli spin depairing is studied in the parallel field case in order to examine an effect of gap nodes on the stability of a Fulde-Ferrell-Larkin-Ovchinnikov (FFLO)-like vortex lattice. We find through a heuristic argument and a model calculation with a four-fold anisotropic Fermi surface that the FFLO-like state is relatively suppressed as the field approaches a nodal direction. When taking account of a couple of experimental results altogether, the present result strongly suggests that the pairing symmetry of CeCoIn$_5$ should be of $d_{xy}$-type. 

\end{abstract}


\maketitle

\section{}

In a recent paper \cite{Ada1} (denoted as I hereafter), we have examined the vortex phase diagram of quasi two-dimensional (Q2D) type II superconductors with strong Pauli paramagnetic (spin) depairing by focusing on ${\bf H} \parallel c$ case with a field ${\bf H}$ perpendicular to the superconducting layers. In contrast to earlier works \cite{GG,Buzdin} taking account of both the orbital and spin depairing effects of the magnetic field in clean limit, the orbital depairing has been incorporated fully and {\it nonperturbatively} there \cite{Ada1}, and two new results opposite to those suggested previously \cite{GG,Buzdin} were found. First of all, the mean field (MF) transition at the $H_{c2}(T)$-line changes from the familiar second-order one to a first-order (MF-FOT) \cite{Izawa,Tayama1,Bianchi1} one at a {\it higher} temperature $T^*$ than a region in which a Fulde-Ferrell-Larkin-Ovchinnikov (FFLO)-like \cite{FF,LO} modulated {\it vortex lattice} may appear. This feature is consistent with data of CeCoIn$_5$ in ${\bf H} \parallel c$ \cite{Izawa,Tayama1,Bianchi1,Bianchi2}. Second of all, a {\it second order} transition curve $H_{\rm FFLO}(T)$ between such a FFLO-like and ordinary vortex lattices remarkably {\it decreases} upon cooling. Interestingly, these two results are also consistent with more recent data of CeCoIn$_5$, suggesting a structural transition to a FFLO state, in ${\bf H} \perp c$ \cite{Radovan,Bianchi2,Matsuda,Tayama2}. A recent ultrasound measurement \cite{Matsuda} also shows that the suggested FFLO state is, as we have argued in I, a kind of vortex lattice. However, it should be further examined theoretically whether this qualitative agreement with the data in ${\bf H} \perp c$ is justified or not. 

In this paper, results of an application of analysis in I to a model for the ${\bf H} \perp c$ case are reported. By including the contributions, neglected in previous works \cite{Ada1,Buzdin}, from the nonGaussian ($|\Delta({\bf r})|^4$ and $|\Delta({\bf r})|^6$) terms of the Ginzburg-Landau (GL) free energy to the spatial gradient parallel to ${\bf H}$, where $\Delta({\bf r})$ is the pair-field, we find that the relative position between $T^*$ and $H_{\rm FFLO}$-line is qualitatively the same as in ${\bf H} \parallel c$ case \cite{Ada1} as far as a spin depairing strength realistic in bulk superconductors is used, and that, at least close to $H_{\rm FFLO}$, the LO state \cite{Buzdin,LO} with periodic nodal planes perpendicular to ${\bf H}$ of $|\Delta|$ is more stable than the FF state \cite{Buzdin,FF} composed of a phase-modulation with keeping $|\Delta|$ fixed. 

A special attention is paid in this paper to a noticeable in-plane angular dependence of the FFLO curve $H_{\rm FFLO}(T)$ found in specific heat \cite{Bianchi2} and magnetization \cite{Tayama2} data of CeCoIn$_5$: The observed FFLO curve in $H \parallel$ [110] lies at higher temperatures than that in $H \parallel$ [100]. This $H_{\rm FFLO}$-anisotropy is much more remakable \cite{Bianchi2} than that of $H_{c2}(T)$ and may give a decisive information on the four-fold anisotropy of the gap function. As far as an {\it in-plane} Fermi velocity anisotropy is negligible, it is heuristically predicted through the following simple argument how a gap anisotropy results in a $H_{\rm FFLO}$-anisotropy: Near the gap nodes where the superconducting gap $\Delta_{\bf k}$ is small, the coherence length $\xi_{\bf k} \simeq \hbar v_{\rm F}/\Delta_{\bf k}$ defined locally in the ${\bf k}$-space is longer \cite{Leggett}. An orbital limiting field $H_{\rm orb}(0)$ is inversely proportional to the square of an averaged coherence length in the plane perpendicular to ${\bf H}$ and hence, is minimal when ${\bf H}$ is directed along the four-fold symmetric gap nodes (or minima). Since a higher $H_{\rm orb}$ will lead to a relatively stronger effect of spin depairing, the FFLO curve and $T^*$, induced by the spin depairing, are expected to lie at higher temperatures as ${\bf H}$ is located along a gap maximum. If comparing the expected $H_{\rm FFLO}$-anisotropy with the observations \cite{Bianchi2,Tayama2} in CeCoIn$_5$, we inevitably reach the conclusion that, in agreement not with the original argument \cite{Izawa} favoring a $d_{x^2-y^2}$-pairing just like in high-$T_c$ cuprates but with a recent report on low $H$ specific heat data \cite{Aoki}, a node (or a minimum) of the gap function of CeCoIn$_5$ is located along the [100]-direction, implying a $d_{xy}$-pairing state. Below, we will show how this conclusion is reinforced through a microscopic derivation of $H_{\rm FFLO}(T)$ taking account of a possible in-plane four-fold anisotropy of the Fermi surface (FS). The present result might require a serious change in the picture on the pairing mechanism of CeCoIn$_5$ based upon similarities on the normal state properties, including the presence of antiferromagnetic fluctuation, to the high $T_c$ 
cuprates \cite{Nakajima}. 

First, let us sketch the outline of MF analysis \cite{Ada1} for ${\bf H} \parallel c$. Throughout this paper, we assume ${\bf H} = H {\hat x}$ and the $d$-wave gap function $w_\phi=\sqrt{2} {\rm cos}(2\phi)$ or $\sqrt{2} {\rm sin}(2\phi)$, where $\phi$ is the azimuthal angle in the $a$-$b$ plane. Within the lowest ($N=0$) Landau level (LL), the GL free energy density in the MF approximation takes the form 
\begin{eqnarray}
{\cal F}_{\rm MF} &=& N(0) \Biggl[ a_0(Q) \langle |\Delta_Q^{(0)}|^2 \rangle + \frac{V_4(Q)}{2} \langle |\Delta_Q^{(0)}|^4 \rangle \nonumber \\
&+& \frac{V_6(Q)}{3} \langle |\Delta_Q^{(0)}|^6 \rangle \Biggr] 
\simeq c_0 + c_2 Q^2 + c_4 Q^4. 
\end{eqnarray}
The essential part of the MF analysis in I is to derive the coefficients, $a_0$, $V_4$, $V_6$, $c_2$, and $c_4$ by starting from the weak-coupling BCS model with a Zeeman (Pauli-paramagnetic) term. 
Here, $N(0)$ is an averaged density of states (DOS) at the Fermi level, and $\langle \,\,\, \rangle$ is the spatial average on $y$ and $z$. $\Delta({\bf r})$ was expanded in terms of the LLs as $\Delta({\bf r}) = \sum_{N \geq 0} \Delta^{(N)}_Q(y,z) u_Q(x)$, and the higher LLs were neglected above. For the LO (FF) state, $u_Q(x)$ takes the form ${\rm cos}(Q x)$ ($ {\rm exp}({\rm i}Q x)$). A Q2D FS with a circular form in $y$-$z$ plane was assumed, although an in-plane anisotropy will be conveniently included as $\phi$-dependences of the Fermi velocity and DOS (See eq.(5) below). For an example, $a_0(Q)$ is, after performing ${\bf k}$-integrals and introducing a parameter integral, expressed by 
\begin{eqnarray}
N(0) a_0(Q)&=& \Biggl\langle \, u^*_Q(x) \biggl(\frac{1}{|g|} - 2 \pi T \nonumber \\
\times \int_0^\infty &d\rho& \frac{{\rm cos}(2 \mu_0 H \rho)}{{\rm sinh}(2 \pi T \rho)}  g^{(0)}(\rho, -{\rm i}\partial_x) \biggr) u_Q(x) \, \Biggr\rangle_x, 
\end{eqnarray}
where $\langle \, \, \, \rangle_x$ denotes the spatial average on $x$, $\mu_0 H$ is the Zeeman energy, and $N(0) |g|$ is the dimensionless 
pairing interaction strength. 
The function $g^{(0)}(\rho, -{\rm i}\partial_x)$ has the form 
\begin{eqnarray}
g^{(0)}(\rho, -{\rm i}\partial_x) &=& N(0) \, 
{\rm exp}(-\rho^2 v_F^2/4 r_H^2) {\rm cos}(-{\rm i} \rho v_F \partial_x), 
\end{eqnarray}
where $r_H$ is the magnetic length, and $v_F$ the Fermi velocity. The extension, $a_N$, of $a_0$ to the $N$-th LL is given by multiplying eq.(3) by ${\cal L}_N(\rho^2v_F^2/2 r_H^2)$, if keeping just terms diagonal with respect to the LLs, where ${\cal L}_N(x)$ the $N$-th Laguerre polynomial. 
The coefficients $V_4(Q)$ and $V_6(Q)$ are derived in a similar manner to above. The coefficients $c_2$ and $c_4$ arise from the $Q$-dependences of $a_0$, $V_4$, and $V_6$. 

The onset $T^*$ of MF-FOT at $H_{c2}$ is determined by $V_4(0)=0$ irrespective of the details of higher order nonGaussian terms of the GL free energy, while $H_{\rm FFLO}(T)$ is defined by $c_2=0$ under the condition $c_4 > 0$. We have verified that the latter condition is always satisfied throughout the computations in the present work, so that the resulting $H_{\rm FFLO}(T)$ is a second order transition line. If the effective strength of spin depairing $\mu_0 H_{\rm orb}^{2D}/(2 \pi k_{\rm B} T_{c0})$ is of order unity or larger, a phase diagram derived numerically in this manner includes a $H_{\rm FFLO}(T)$-line decreasing upon cooling, where $\mu_0 H$ is the Zeeman energy, and $H_{\rm orb}^{2D}$ is the orbital limiting field in 2D limit. 
In Ref.1 where the $V_4$ and $V_6$-contributions to $c_2$ were neglected, the LO and FF states had the same $H_{\rm FFLO}$-line, while we find that the instability of the straight vortex lattice into the LO vortex state \cite{Buzdin,LO} occurs at a slightly higher tempeature than that into the FF state \cite{Buzdin,FF}. Hence, at least close to $H_{\rm FFLO}(T)$, the LO state becomes the ground state in $H_{\rm FFLO} < H < H_{c2}$. Further, we find that the $V_6$-contribution to $c_2$ is quantitatively negligible, while the $H_{\rm FFLO}$-line is pushed by the corresponding $V_4$-contribution down to a lower temperature region in which $H_{c2}$ and the vortex state just below it are described by the $N=1$ LL. Thus, at least within the weak-coupling BCS model, a FFLO state in ${\bf H} \parallel c$ rarely occurs because such a $N=1$ LL vortex lattice has no FFLO-like modulation \cite{Ada1}. We guess that a slight specific heat anomaly \cite{Bianchi2} in CeCoIn$_5$ in ${\bf H} \parallel c$ at low enough temperatures may be rather due to a transition between straight vortex lattices in $N=0$ and $N=1$ LLs. A detailed study of this transition into a $N=1$ LL state will be reported elsewhere. 

Now, let us turn to the ${\bf H} \perp c$ case. Although, in principle, the above analysis can be extended to a Q2D system with a cylindrical FS under ${\bf H}$ perpendicular to the cylindrical axis, we have chosen to work in an elliptic FS elongated along $z$($\parallel c$)-axis and with the dispersion relation 
$\varepsilon_k=\hbar^2 \sum_{j=x,y,z} \gamma_j^{-2} k_j^2/(2 \overline{m})$ 
under ${\bf H} \parallel {\hat x}$ in order to make numerical calculations more tractable, where $\gamma_x = \gamma_y = \gamma^{-1/2}$, and $\gamma_z 
= \gamma$ 
with $\gamma \geq 1$ and a constant $\overline{m}$. 
We expect the case with a moderately large $\gamma$-value to {\it qualitatively} describe essential features in the realistic Q2D case. 
By isotropize the ${\bf k}$ vector as 
$k_j = \gamma_j k_{\rm F} {\hat r}_j$, where ${\bf {\hat r}} =  ({\rm cos}\phi {\rm sin}\theta, {\rm sin}\phi {\rm sin}\theta, {\rm cos}\theta)$ is the unit vector in the spherical coordinate, the velocity ${\bf v}$ on FS is written as $v_{j} = \gamma_j^{-1} v_{\rm F} {\hat r}_j$. A Jacobian $\sqrt{\gamma {\rm sin}^2 \theta + \gamma^{-2} {\rm cos}^2 \theta}$ accompanying the angular integral along FS is exactly cancelled by the angular dependence of 
DOS $N(\theta)= N(0) v_F/\sqrt{\sum_j v_j^2}$. 
Again, the {\it in-plane} (four-fold) anisotropy of FS will be first 
neglected. 
Then, the GL free energy within $N=0$ LL takes the form of eq.(1), and the function $g^{(0)}(\rho,-{\rm i}\partial_x)$ appearing in $a_0(Q)$ (see eq.(3)) is replaced in the present case by 
\begin{eqnarray}
g_\parallel^{(0)}(\rho, -{\rm i}\partial_x) &=& \int \frac{{\rm sin}\theta d\theta d\phi}{4\pi} N(0) |w_\phi|^2 
{\rm exp}(-\rho^2 {\overline v}^2_{yz}/4 r_H^2) \nonumber \\
&\times& {\rm cos}(-{\rm i} \rho v_x \partial_x) , 
\end{eqnarray}
\begin{figure}[t]
\scalebox{0.42}[0.45]{\includegraphics{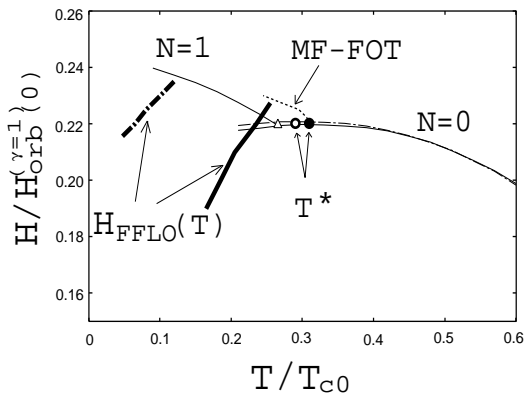}}
\caption{$H$-$T$ mean field phase diagram obtained using $\gamma=3$ and with no in-plane FS anisotropy. The transition or crossover positions in ${\bf H} \parallel$ gap-maximum ($\parallel$ gap-node or minimum) are expressed by the solid curves and filled circle (chain curves and open circle). The dotted curve and open triangle denote, respectively, the MF-FOT line and the position at which the two solid curves $a_N(0)=0$ in $N=0$ and $1$ merge with each other.}
\label{fig: }\end{figure}
where ${\overline v}^2_{yz} = {\tilde \eta} \gamma^{-1} v_y^2 + \gamma {\tilde \eta}^{-1} v_z^2$. The parameter ${\tilde \eta}$ is insensitive to the uniaxial anisotropy $\gamma$ but dependent on $T$ and needs to be determined by maximizing $H_{c2}(T)$. By focusing on the low $T$ region, we find that ${\tilde \eta}$ takes a value between 0.4 and 0.5 depending on the relative angle between ${\bf H}$ and a nearest nodal direction. Using this parameter, the anisotropy in spatial variations of $\Delta({\bf r})$ within the $y$-$z$ plane is given by $\gamma/{\tilde \eta}$. Except for modifications indicated above, the corresponding quartic and 6th order terms of the GL free energy are derived by closely following the analysis in I. We choose $\alpha_\parallel = \mu_0 H_{\rm orb}^{(\gamma=1)}(0)/k_{\rm B} T_{c0}$ as a measure of the spin depairing strength in ${\bf H} \perp c$, where $H_{\rm orb}^{(\gamma=1)}(0)$ is the orbital limiting field in the isotropic case. 

In Fig.1, the resulting phase diagram is shown to illustrate how the $H_{\rm FFLO}(T)$-position depend upon the relative angle between ${\bf H}$ and the nodal directions. Thin solid (chain) curves are defined by $a_N(0)=0$, and the $H_{c2}(T)$ in $T > T^*$ in each case is given by each $a_0(0)=0$ line. In agreement with the heuristic argument given earlier, $H_{\rm FFLO}(T)$ and $T^*$ are shifted to higher temperatures as the in-plane field is directed along a gap-maximum, reflecting an enhanced spin depairing in this field configuration. As in ${\bf H} \parallel c$, the FFLO state at least close to $H_{\rm FFLO}$ has the LO-like variation. We have verified by combining our numerical calculations with an analytical calculation with the orbital depairing perturbatively included that such an in-plane $H_{\rm FFLO}$-anisotropy is absent without the orbital depairing (i.e., when $\alpha_\parallel=\infty$) and {\it monotonously} increases with decreasing $\alpha_\parallel$. 
In contrast, it is not easy to properly predict the corresponding anisotropy (in-plane angular dependence) of $H_{c2}(T)$-curve: First, the depression of $H_{c2}$ due to the spin depairing is larger as the corresponding $H_{\rm orb}(0)$ is higher, and hence, the $H_{c2}$-magnitude may not have a monotonous $\alpha_\parallel$-dependence. Second, the MF-FOT line of $H_{c2}$ is directly determined by the details of the nonGaussian terms other than the quartic one in GL free energy \cite{Ada1} and hence, is quantitatively affected by our assumption of keeping the nonGaussian terms only up to $|\Delta|^6$ in eq.(1). Actually, the rapid increase of MF-FOT line on cooling {\it just} below $T^*$ arises due to an extremely small $V_6(0)$ near $T^*$ and might flatten if we can numerically include the $|\Delta|^8$ and higher order terms. In contrast, the $V_6$-contribution to $c_2$ (i.e., to $H_{\rm FFLO}(T)$) was negligible, like in ${\bf H} \parallel c$ case, consistently with the smallness of $V_6(0)$ mentioned above. We expect that the $H_{\rm FFLO}(T)$ curve is less sensitive to the neglect of the $|\Delta|^8$ and higher order GL terms. For these reasons, we will focus hereafter on $T^*$ and $H_{\rm FFLO}$ which directly measure the (effective) 
spin depairing strength. 
The resulting anisotropies of $T^*$ and $H_{\rm FFLO}$ in Fig.1 qualitatively agree with those of CeCoIn$_5$ in ${\bf H} \perp c$ \cite{Bianchi2,Tayama2} if a gap node (or minimum) is located along [100]. As already mentioned, the MF-FOT line in $N=0$ LL needs to lie above the corresponding $a_1(0)=0$ line in order for $H_{\rm FFLO}(T)$ to be realized as a transition line. As Fig.1 shows, this condition manages to be satisfied in contrast to the ${\bf H} \parallel c$ case. 

In order to examine how the result in Fig.1 is affected by the {\it in-plane} FS anisotropy, let us next introduce it as a Fermi velocity anisotropy 
in a similar manner to Ref.\cite{Nakai} like 
\begin{eqnarray}
v_{\rm F} \,\,\, &\to& \,\,\, v_{\rm F}(\phi) = v_{\rm F} (1 + \beta {\rm cos}(4\phi)),  
\end{eqnarray}
\begin{figure}[t]
\scalebox{0.40}[0.42]{\includegraphics{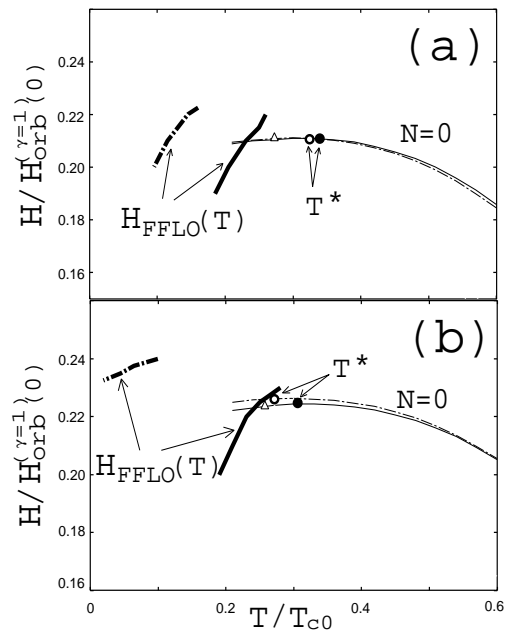}}
\caption{Results corresponding to Fig.1 in the cases (a) ($|\beta|=0.2$) and (b) ($|\beta|=0.1$) defined in the text.}
\label{fig:}\end{figure}
where $|\beta| < 1$, accompanied by the replacement $
N(0) \,\, \to \,\, N(0) v_{\rm F}/v_{\rm F}(\phi)$ 
in any angular integral (see eq.(4)). Except these replacements in our calculation, the derivation of phase diagrams is quite the same as that of Fig.1. When $\beta > 0$ ($< 0$), the Fermi velocity becomes maximal (minimal) along ${\hat x}$. By combining these two cases with the two candidates, $\sqrt{2} {\rm cos}(2\phi)$ and 
$\sqrt{2} {\rm sin}(2\phi)$, of $w_\phi$, we have four different cases of the relative anisotropies under a fixed ${\bf H} \parallel {\hat x}$. We will classify them into two categories, (a) $w_\phi=\sqrt{2}{\rm cos}(2\phi)$ with $\beta < 0$ and $w_\phi=\sqrt{2} {\rm sin}(2\phi)$ with $\beta > 0$, and (b) $w_\phi=\sqrt{2}{\rm cos}(2\phi)$ with $\beta > 0$ and $w_\phi=\sqrt{2} {\rm sin}(2\phi)$ with $\beta < 0$. This classification is motivated by the result \cite{Nakai} that, in the category (a), the Fermi velocity anisotropy and the pairing anisotropy favor two different orientations, competing with each other, of the square vortex lattices to be realized in four-fold anisotropic $d$-wave superconductors in ${\bf H} \parallel c$, while such a competition does not occur in (b). In Fig.2, the resulting phase diagrams for the categories (a) and (b) are given. In the case (a), the angular dependences of $H_{\rm FFLO}$ and $T^*$ are weakened by the FS anisotropy compared with those in Fig.1, while the opposite tendency is seen in the case (b). This result can be understood as follows by noting that the orbital depairing strength local in the ${\bf k}$-space is measured in the present case by $v_y^2$ in eq.(4) (Note that, in 2D limit, $v_z^2$ is absent there). By focusing on the case with ${\bf H}$ parallel to a gap node and noting $|w_\phi|^2$ in the integrand of eq.(4), one will notice that a nonzero $|\beta|$ tends to increase (decrease) contributions of $v_y^2$, on average, when $\beta < 0$ ($\beta > 0$). Thus, an enhanced orbital depairing in ${\bf H}$ parallel to a node of the case (b) additionally reduces $H_{\rm FFLO}$ so that the difference between the two cases in Fig.2 follows. Bearing in mind the general character of this interpretation, we believe that the results in Fig.2 are not qualitatively changed by a refinement of microscopic description. 

The above results commonly show a $H_{\rm FFLO}(T)$-line shifting to higher temperatures as the in-plane field approaches a gap-maximum and, compared with the data in CeCoIn$_5$ \cite{Bianchi2,Tayama2}, imply a $d_{xy}$-state as the pairing state of this material. Although one might consider a possibility of $d_{x^2-y^2}$-pairing based on the fact that an extremely strong FS anisotropy in the case (a) may reverse the anisotropies of $T^*$ and $H_{\rm FFLO}$, such a strong FS anisotropy of the case (a) should result \cite{Nakai} in a square vortex lattice with an orientation due to the FS anisotropy and hence, contradicts not only the specific heat data \cite{Aoki} but the observed orientation \cite{square} of ${\bf H} \parallel c$ square vortex lattice. Therefore, an inclusion of FS anisotropy {\it reinforces} our conclusion favoring a $d_{xy}$-pairing, although a moderate FS anisotropy competitive in ${\bf H} \parallel c$ with the gap anisotropy (i.e., of the case (a)) is needed for quantitative understandings. 

In conclusion, the mean field phase diagram of a type II superconductor with strong Pauli paramagnetic depairing and with a four-fold symmetric $d$-wave pairing was qualitatively studied in the parallel field case. The region in which the FFLO vortex phase appears is enlarged when the in-plane field is directed along a gap-maximum. This result is reinforced by including in-plane FS anisotropies and strongly suggests a $d_{xy}$ pairing as the best candidate of the gap function of CeCoIn$_5$ in spite of the electronic similarities \cite{Nakajima} to that of high $T_c$ cuprates. A reinterpretation of thermal conductivity data by Izawa et al. \cite{Izawa} can be seen in Ref.\cite{Aoki}. The present theory should be applicable to examining the pairing state of other materials, such as an organic material \cite{Izawa2,Tanatar}, showing a remarkable Pauli paramagnetic depairing. 

We thank Y. Matsuda, K. Machida, T. Sakakibara, and T. Tayama for informative discussions. The present work was finantially supported by a Grant-in-Aid from the Ministry of Education, Culture, Sports, Science, and Technology, Japan.


\section{References}

\begin{thebibliography}{99}
\bibitem{Ada1} H. Adachi and R. Ikeda, Phys. Rev. B {\bf 68}, 184510 (2003). 
\bibitem{GG} L.W. Gruenberg and L. Gunther, Phys. Rev. Lett. {\bf 16}, 966 
(1966). 
\bibitem{Buzdin} M. Houzet and A. Buzdin, Phys. Rev. B {\bf 63}, 184521 
(2001). 
\bibitem{Izawa} K.Izawa, H.Yamaguchi, Y.Matsuda, H.Shishido, R.Settai, and 
Y.Onuki, Phys. Rev. Lett. {\bf 87}, 057002 (2001). 
\bibitem{Tayama1} T.Tayama, A.Harita, T.Sakakibara, Y.Haga, H.Shishido, 
R.Settai,and Y.Onuki, Phys. Rev. B {\bf 65}, 180504 (2002). 
\bibitem{Bianchi1} A.Bianchi, R.Movshovich, N.Oeschler, P.Gegenwart, 
F.Steglich, J.D.Thompson, P.G.Pagliuso, and J.L.Sarrao, Phys. Rev. Lett. {\bf 89}, 137002 (2002). 
\bibitem{FF} P. Fulde and R. A. Ferrell, Phys. Rev. {\bf 135}, A550 (1964). 
\bibitem{LO} A. I. Larkin and Y. N. Ovchinnikov, Sov. Phys. JETP {\bf 20}, 762 (1965). 
\bibitem{Bianchi2} A.Bianchi, R.Movshovich, C.Capan, P.G.Pagliuso, and J.L.Sarrao, Phys. Rev. Lett. {\bf 91}, 187004 (2003). 
\bibitem{Radovan} H.A. Radovan, N.A. Fortune,T.P.Murphy, S.T.Hannahs, E.C.Palm, S.W.Tozer, and D.Hall, Nature {\bf 425}, 51 (2003). 
\bibitem{Matsuda} T.Watanabe et al., cond-mat/0312062. 
\bibitem{Tayama2} T. Tayama, private communication. 
\bibitem{Leggett} I. Kosztin and A. J. Leggett, Phys. Rev. Lett. {\bf 79}, 135 (1997). 
\bibitem{Aoki} H.Aoki, T.Sakakibara, H.Shishido, R.Settai, Y.Onuki, P.Miranovic, and K.Machida, J. Phys. Condens. Matter {\bf 16}, L13 (2004). 
\bibitem{Nakajima} V.A.Sidorov, M.Nicklas, P.G.Pagliuso, J.L.Sarrao, Y.Bang, A.V.Balatsky, and J.D.Thompson, Phys. Rev. Lett. {\bf 89}, 157004 (2002). 
\bibitem{Nakai} N.Nakai, P.Miranovic, M.Ichioka, and K.Machida, Phys. Rev. Lett. {\bf 89}, 237004 (2002). 
\bibitem{square} M.R. Eskildsen, C.D.Dewhurst, B.W.Hoogenboom, C.Petrovic, and P.C.Canfield, Phys. Rev. Lett. {\bf 90}, 
187001 (2003). 
\bibitem{Tanatar} M.A.Tanatar, T.Ishiguro, H.Tanaka, and H.Kobayashi, Phys. Rev. B {\bf 66}, 134503 (2002). Their identification of FFLO transition line seems to be justified by a similar result of recent thermal conductivity data [C.Capan et al., cond-mat/0401199] in CeCoIn$_5$, if the fluctuation in the organic material is strong enough to erase its discontinuous behavior at $H_{c2}(T)$ at low $T$ \cite{Ada1}. 
\bibitem{Izawa2} K.Izawa, H.Yamaguchi, T.Sasaki, and Y. Matsuda, Phys. Rev. Lett. {\bf 88}, 027002 (2002). 

\end{thebibliography}
\end{document}